\documentclass[manuscript]{aastex}

\slugcomment{}

\shorttitle{Collapsed Cores in Globular Clusters}
\shortauthors{Djorgovski et al.}

\begin{document}


\title{Diagnosis of Magnetic and Electric Fields	\\
of Chromospheric Jets through Spectropolarimetric	\\
Observations of \ion{H}{1} Paschen Lines}


\author{T. Anan}
\affil{Kwasan and Hida Observatories, Kyoto University,
    Gifu, Japan, 506-1314}
\email{anan@kwasan.kyoto-u.ac.jp}

\author{R. Casini}
\affil{High Altitude Observatory, National Center for Atmospheric 
	Research,\altaffilmark{1} Boulder, CO 80301}
\email{casini@ucar.edu}

\and

\author{K. Ichimoto}
\affil{Kwasan and Hida Observatories, Kyoto University,
    Gifu, Japan, 506-1314}
\email{ichimoto@kwasan.kyoto-u.ac.jp}


\altaffiltext{1}{The National Center for Atmospheric Research is
sponsored by the National Science Foundation}


\begin{abstract}
Magnetic fields govern the plasma dynamics in the outer layers of the 
solar atmosphere, and electric fields acting on neutral atoms 
that move across the magnetic field enable us to study the dynamical coupling 
between neutrals and ions in the plasma.
In order to measure the magnetic and electric fields of chromospheric jets, 
the full Stokes spectra of the Paschen series of neutral hydrogen in a 
surge and in some active region jets that took place at the solar limb were 
observed on May 5, 2012, using the spectropolarimeter of the Domeless 
Solar Telescope at Hida observatory, Japan. First, we inverted the Stokes 
spectra taking into account only the effect of magnetic fields on the 
energy structure and polarization of the hydrogen levels. Having found 
no definitive evidence of the effects of electric fields in the observed 
Stokes profiles, we then estimated an upper bound for these fields by calculating the polarization degree 
under the magnetic field configuration derived in the first step, with 
the additional presence of
a perpendicular (Lorentz type) electric field of varying strength.
The inferred direction of the magnetic field on the plane of the sky
(POS)
approximately aligns to the active region jets and the surge, with magnetic 
field strengths in the range $10\,\mathrm{G} < B < 640 \,\mathrm{G}$ for 
the surge.
Using magnetic field strengths of $70$, $200$, and $600\,\mathrm{G}$, we 
obtained upper limits for possible electric fields of $0.04$, $0.3$, and 
$0.8\,\mathrm{V\,cm^{-1}}$, respectively.
This upper bound is conservative, since in our modeling we neglected 
the possible contribution of collisional depolarization.
Because the velocity of neutral atoms of hydrogen moving across the 
magnetic field derived from these upper limits of the Lorentz electric 
field is far below the bulk velocity of the plasma perpendicular to the 
magnetic field as measured by the Doppler shift, we conclude 
that the neutral atoms must be highly frozen to the magnetic field in 
the surge.
 

\end{abstract}





\section{Introduction}
\label{sec.intro}

Magnetic fields govern the plasma dynamics in the outer layers of the 
solar atmosphere.
The magnetic energy built up by convection in and below the photosphere 
is transferred from the photosphere into the heliosphere. 
During the transfer of energy into the outer layers of the solar 
atmosphere, i.e., chromosphere and corona, the magnetic energy is 
partially converted into plasma kinetic energy, causing various 
active phenomena in the solar atmosphere. 
Thus the measurement of the magnetic field is of crucial importance to 
identify the mechanisms responsible for the dynamical phenomena in the
solar atmosphere.

Several processes can generate polarization in spectral lines in response to 
the presence of magnetic and electric fields in the radiation emitting plasma.
The Zeeman and Stark effects are produced by the energy separation of the 
atomic levels into multiple sublevels due to the potential energy of the
radiating atom in the external fields. As the splitted components have
different polarization properties, this results in a wavelength dependent 
polarization across the spectral line. In the absence of velocity or
field gradients in the emitting plasma, the broadband (i.e., wavelength
integrated) polarization from the Zeeman and Stark effect vanishes.
In contrast, in the presence of \emph{atomic polarization} (i.e., population 
imbalances and quantum interference among the magnetic sublevels; e.g., \citealt{landi04}
), the emitted (scattered) radiation
is typically characterized by a non-vanishing broadband 
polarization. The two dominant characteristics of atomic polarization are atomic 
\emph{alignment} and \emph{orientation}. Atomic alignment is a state of 
population imbalances between magnetic sublevels with different absolute
values of their magnetic quantum numbers. Atomic orientation is 
characterized by population imbalances between sublevels with positive 
and negative magnetic quantum numbers.
The effects of atomic alignment and orientation in the scattered
radiation are observed in the form of broadband linear and circular 
polarization, respectively.
In the process of radiation scattering, the anisotropy of the incident 
radiation field produces atomic alignment, and the scattered light is 
therefore linearly polarized, with a magnitude depending on the 
scattering geometry.
The presence of a magnetic or electric field produces  
a relaxation of the atomic coherence among the magnetic sublevels
resulting in a modification of the polarization (typically, depolarization 
and rotation) of the scattered radiation, known as the Hanle effect.
%
Strong magnetic or electric fields also cause a conversion of atomic 
alignment into atomic orientation when the energy splitting of the
sublevels induced atomic level crossings 
\citep[A-O mechanism, ][]{lehmann64,landi82,kempetal84,favatietal87,casini05b}.
Collisions of the scattering atoms with free electrons and protons 
tend to reduce the atomic level polarization.
For the typical density of the upper solar chromosphere, collisional 
transitions between the fine-structure levels pertaining to the same 
Bohr level of the neutral hydrogen may play a significant role in the 
depolarization process \citep{bommier86,sahal96,stepan11}.

Recent progress in the modeling of scattering polarization \citep{bueno02, ariste02, bueno07} and in the development of increasingly
sensitive spectropolarimetric instrumentation 
\citep[e.g.][]{collados07, jaeggli10, anan12} enables us to measure 
the chromospheric magnetic field by interpreting the observed
polarization in terms of the Zeeman and Hanle effects.
Recently, the magnetic field of some types of chromospheric phenomena 
have been measured, for instance, in filaments \citep{bueno02}, 
prominences \citep{casini03, merenda06, sasso11}, spicules 
\citep{lopez_casini05, bueno05}, 
and in regions of emerging magnetic flux \citep{lagg04}.

Chromospheric jets called ``surges'' often occur near sunspots, 
in the proximity of neutral points of the magnetic fields 
\citep{rust68}, in flux emergence regions \citep{kurokawa88}, or in 
association with magnetic flux cancellation \citep{gaizauskas96}.
They reach heights up to 200 Mm and velocities of 50 to 200 km s$^{-1}$ 
\citep{roy73}.
Recent observations have found tiny chromospheric jets outside sunspots 
in active regions \citep{shibata07, morita10}; their typical length and 
velocity are 1 to 4 Mm and 5 to 20 km s$^{-1}$, respectively 
\citep{nishizuka11}.
In addition, jet-like features are observed in sunspot penumbrae where 
stronger, inclined magnetic fields and weaker, nearly 
horizontal magnetic fields co-exist at small spatial scales
\citep{katsukawa07}.
These jets are interpreted as the result of magnetic energy release 
caused by a change of magnetic field topology 
\citep[e.g.][]{shibata07, ryutova08, nakamura12, takasao13}.
However, there are no studies reporting direct measurements of the 
magnetic field of jets in the chromospheric lines.

Unlike the measurement of magnetic fields, the study of electric fields
in solar plasmas has been given little attention. This is partially
due to the fact that appreciable quasi-static electric fields on
macroscopic spatial scales are unlikely in the solar atmosphere, since 
the relaxation time for the electric charge is very short in the highly 
conductive solar atmosphere. Electric fields generated in magnetic 
reconnection events occur at spatial scales far smaller than 
the spatial resolution of current spectropolarimetric instrumentation.
On the other hand, the spatial scale of Lorentz electric fields associated 
with plasma motions across the magnetic field can be large enough to be 
resolved with existing solar telescopes, and the associated Stark effect
can be detected in a weakly ionized plasma that is able to reach
sufficient bulk velocities across the magnetic field.

Since detection of the electric field experienced by neutral atoms 
moving across the magnetic field can tell us about the degree of 
dynamical coupling between the neutrals and the ionized plasma, the 
observation and detection of these \emph{motional} electric fields is 
of particular interest for understanding the dynamics of partially 
ionized plasmas in the chromosphere.

In order to possibly detect electric fields, we carried out 
spectropolarimetric observations of the high Paschen series 
(i.e., large principal quantum number of the upper level) of 
neutral hydrogen. The choice of hydrogen was motivated because
the linear Stark effect (a first order perturbation of the atomic
level energy) only occurs in hydrogen-like ions, whereas
multi-electron atoms tend to show only quadratic (and higher order)
energy effects in the presence of an external electric field. In the
case of the linear Stark effect, the splitting of the energy levels 
is also roughly proportional to the square of the principal 
quantum number 
\citep[e.g.][]{moran91, foukal95}, so the use of high transitions in
a hydrogen line series is desirable in order to increase the 
magnitude of the
polarization signal. \cite{moran91} and \cite{foukal95} estimated the 
electric field in post flare loops and prominences by studying the line 
width variation of the linear polarization profile between two
orthogonal states of polarization. The implicit assumption in that study
was that atomic polarization of such highly excited initial levels
could 
be neglected since they should mainly be populated by isotropic 
recombination processes \citep{casini96}.
\citet{foukal95} measured a surge that occurred in association with a 
solar flare and obtained a value of the electric field of 
$\sim 35\,\mathrm{V\,cm^{-1}}$.

Formalisms for the description of hydrogen line polarization in the 
simultaneous presence of magnetic and electric 
fields in both local thermodynamic 
equilibrium (LTE) \citep{casini93} and non-LTE including atomic
polarization \citep{casini05b} have been developed.
In LTE, an electric field of the order of 1 V cm${\rm ^{-1}}$ produces 
a linear polarization of the order of 0.1\% through the linear Stark effect.
\citet{casini05b} pointed out that if an electric field is present 
in the non-LTE scattering process, magnetic field strengths of order of 
10 G can produce a significant level of atomic orientation 
through the 
A-O mechanism, comparable to that produced by a magnetic field of the 
order of 10$^3$ G in the absence of an electric field. This is because,
in the presence of an electric field, magnetic-induced level crossing
can happen between $\Delta L=1$ levels that are separated in 
energy by a difference comparable to
the Lamb shift, rather than by the much larger L-S coupling fine
structure energy separation.
In the scattering process with both magnetic and electric fields, an 
electric field of the order of 0.1 V cm${\rm ^{-1}}$ is sufficient to
modify the linear polarization of the observed Paschen 
lines by about 1\%.

\citet{gilbert02} evaluated the drain speed of neutral atoms across 
the magnetic field in a simple prominence model with a partially 
ionized plasma, in which the solar gravitational force balances with 
the frictional force proportional to the relative flow of the neutral 
and ionized components.
Their analytical calculations in steady state show the downflow 
velocity of neutral hydrogen to be $\sim 1.6\,\mathrm{km\,s^{-1}}$ 
for a density of $10^{10} \,\mathrm{cm^{-3}}$, assuming a ionization 
fraction of hydrogen of $10^{-3}$ and a ionization fraction of 
helium of $10^{-2}$.
If we assume a magnetic field strength of
$100\,\mathrm{G}$, 
the electric field experienced by neutral atoms moving across the 
magnetic field is $\sim 0.16\,\mathrm{V\,cm^{-1}}$, and it is possible 
that the electric field modifies the polarization of the 
scattered radiation.
In dynamic phenomena, electric fields can be stronger than in the 
steady state case. For instance, the neutral hydrogen in a surge is 
ejected with a velocity of 50 to 200 km s$^{-1}$ with a whiplike motion 
\citep{roy73, nishizuka08}, while neutral hydrogen in 
prominences descends and rises along vertical threads with a 
velocity of 5 to 20 km s$^{-1}$ \citep{zirker98,berger08}.

In order to study the magnetic and electric fields in chromospheric 
jets, we observed the full Stokes spectra of the Paschen series of 
\ion{H}{1} in active region jets that took place on the solar limb 
on May 5, 2012. For convenience, throughout this paper
we use the two terms ``surge'' and ``jet'' 
to refer to large and small events, respectively. In the 
following sections, we describe the details of the observations 
(Sec. \ref{sec.obs}), the inference of the magnetic and
possible electric 
fields (Sec. \ref{sec.res}), and finally we discuss our measurements and 
provide our conclusions (Sec. \ref{sec.dis_sum}).

\section{Observations and data reduction}
\label{sec.obs}

Several jets in the active region NOAA 11476 were observed in 
several \ion{H}{1} Paschen lines using the universal spectropolarimeter 
\citep{anan12} of the Domeless Solar Telescope \citep{nakai85} at Hida 
observatory, Japan.
The polarimeter uses a continuously rotating achromatic waveplate as 
polarization modulator and provides the full Stokes vector of any 
spectral regions between 4000\,\AA\ and 11000\,\AA\ for all the 
spatial points along the spectrograph slit, with a spatial sampling of 
$\sim 0.4$\,arcsec/px. 
The full Stokes spectra so obtained were calibrated for instrumental 
polarization using a predetermined Mueller matrix of the telescope. 
The expected error on the determination of a given Stokes polarization
parameter $S=Q,U,V$, can be expressed as \citep{ichimoto08,anan12}
\begin{equation}
\delta \Biggl( \frac{S}{I} \Biggr)
	= a\,\frac{S}{I} + \sigma
\end{equation}
where $a$ is the \emph{polarimetric accuracy} of the calibration 
and $\sigma$ is the \emph{rms polarimetric sensitivity} of the
observation, which is measured by the statistical noise with respect 
to the peak intensity of the spectral line. 
Our calibration procedure guarantees that $|a|\lesssim 0.05$. Estimates 
of $\sigma$ for the different Paschen lines in our observations 
are given in the next paragraph.

The observation ran from 2:32\,UT to 3:50\,UT on 2012 May 5.
Two beams with orthogonal polarizations are taken simultaneously with a CCD camera (Prosilica GE1650) with a spectral sampling of 70\,m\AA/pix and an exposure time of 500 msec, and 99 frames are integrated in 49.5 sec, while the spin rate of the rotating waveplate is 0.1 rev./sec.
Full Stokes spectra were observed in ``sit-and-stare'' mode for all 
P$n$ lines of the Paschen series of \ion{H}{1}, for the values 
$n=7,9,10,11,12,13,15,18,19,20$ of the principal quantum number of the 
upper level. These lines were observed sequentially by rotating 
the spectrograph grating.
The rms polarization sensitivity for the strongest 
lines P7, P9, P10, P11, P12, and P13, was estimated to be, respectively, 
$\sim 2\times 10^{-3}$, $2\times 10^{-3}$, $3\times 10^{-3}$, 
$3\times 10^{-3}$, $3\times 10^{-3}$, and $4\times 10^{-3}$.
We did not analyze the Stokes spectra for the Paschen lines above P15, 
because the observed signals were lower than 10\% of the scattered
continuum outside the solar limb, and were thus dominated by flat field
errors. 
The line wavelengths were determined by identifying neighboring 
absorption lines in the background scattered spectrum in the solar 
atlas of \citet{kurucz84}.
Because motion of the solar image on the slit caused by the seeing 
and telescope guiding error was approximately $1$ arcsec in amplitude 
during the run of the observation, we averaged the Stokes spectra in 
the spatial direction over 2\,arcsec. 
The properties of the observed data set are summarized in Table 1.

The slit was placed outside the solar limb, approximately parallel to it, 
above active region NOAA 11476.
The slit width and length were 1.28\,arcsec and 128\,arcsec, respectively.
In our analysis, the slit identifies the reference direction of
polarization, along which Stokes $Q$ is defined to be positive.

Figure~\ref{fig.obs} shows examples of the observed Stokes spectra of the 
P7, P9, P10, P11, P12, P13, and P15 lines with slit-jaw images taken at
the H$\alpha$ line center.
In the spectral ranges of P13 (8665\,\AA) and P15 (8545\,\AA) there are 
strong emission lines of \ion{Ca}{2} 8662\,\AA\ and \ion{Ca}{2} 8542\,\AA, 
whose peaks are strongly saturated. 
During the observation, some jets and a surge took place across the slit.   
The jets were observed in P7 between 2:32\,UT and 2:47\,UT, while the 
surge was observed in the P9, P10, P11, P12, P13, P15, P18, P19, and P20 
during the second half of the observation period.
The distance of the slit to the visible limb for the observation 
of the jets and the surge was, respectively, $\sim 10$\,arcsec and 
$\sim 15$\,arcsec.

After correcting our data for instrumental polarization, the 
Stokes spectra still show a residual polarization signal in the continuum 
around the Paschen lines. This is likely caused by stray light in the 
instrument, and indicates that the line signal in the jets and surge 
is also affected by this spurious polarization. 
We simply corrected for it by subtracting the polarization offset 
that is obtained by averaging the Stokes signal in spatial positions along the slit outside the detectable emissions of Paschen lines.
These signals are typically dominated by the scattered photospheric spectrum, 
showing evidence of absorption features in the spectral region of 
interest. However, in the stray light spectra observed in our data, 
the photospheric signal of the Paschen lines is not distinguishable.

We also cannot exclude that part of the spurious polarization 
offset may be due to residual polarization cross-talk from Stokes $I$ 
after the polarization calibration. In particular, this could be
caused by imperfect spatial coregistration and/or intensity rescaling of 
the two beams with orthogonal polarizations before subtraction. Some
evidence of this can be glimpsed from the Stokes $Q$, $U$, and $V$ spectra 
of Figure~1, as some of these spectra clearly show a spurious signal in
correspondence with the slit-jaw hairlines.
We take into account this 
possibility in Sec.~\ref{sec.res.elec}, when we estimate the effects 
of possible Lorentz fields on the polarization of the observed lines. 

\section{Diagnosis of magnetic and electric fields}
\label{sec.res}

\begin{figure}
\begin{center}
\includegraphics[angle=0,scale=1.0,width=140mm]{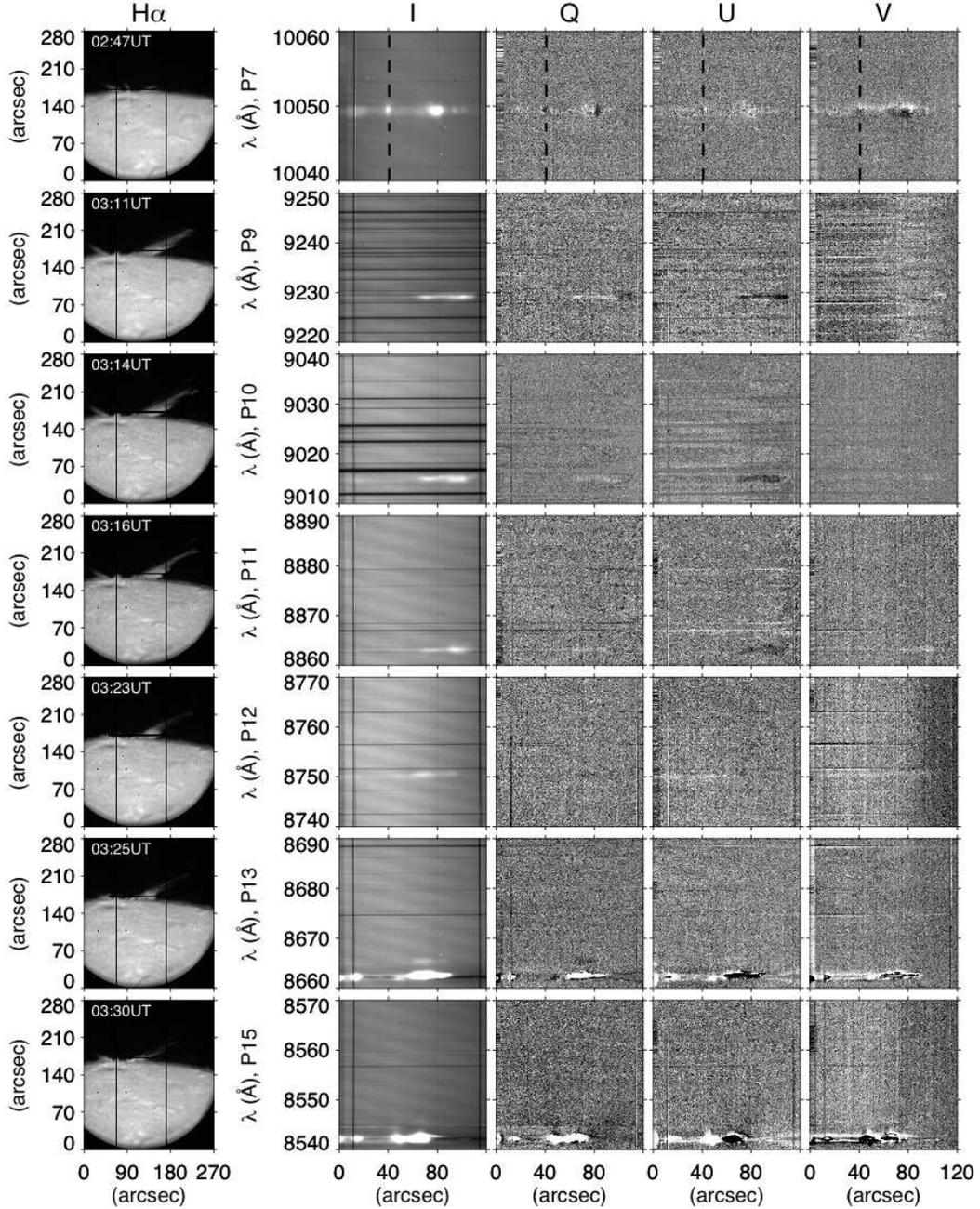}
\end{center}
\caption{
Observed Stokes spectra of the Paschen series of neutral 
hydrogen (4 right columns) and slit-jaw images in H$\alpha$ (left column).
The data set in each row were obtained simultaneously.
The horizontal lines in the slit-jaw images show the spectrograph slit.
The two vertical lines in the slit-jaw images and the $I$ spectral 
images show the spectrograph hairlines. 
The vertical dashed lines in the Stokes spectra of the P7 
denote the location of the profiles shown in Figure \ref{fig.fit}. 
The strong emissions features in the frames of P13 and P15 are 
the \ion{Ca}{2} 8662\AA\ and \ion{Ca}{2} 8542\AA\ lines, respectively.
		}
\label{fig.obs}
\end{figure}

Figure \ref{fig.obs} and \ref{fig.fit} show the observed Stokes spectra of the Paschen series of \ion{H}{1} and an example of Stokes profiles of P7, respectively.
It must be noted that the shape of the linear polarization signals resemble
that of the intensity profile. This suggests the presence of atomic 
polarization in the upper levels of the transitions of neutral hydrogen in 
the studied jets and surge, and puts into question the assumption that
was made in previous studies of electric field measurements by 
spectro-polarimetric observations, that the atomic polarization of highly 
excited levels should be negligible \citep[e.g.][]{foukal91,casini96},
presuming that these levels are mostly populated by electron
recombination.
\emph{Our observations indicate instead that optically pumped atomic
polarization must be significant, at least for the upper levels 
of the Paschen lines that we considered.}

\citet{casini05b} derived a formalism for modeling the scattering 
polarization of hydrogen lines in the presence of both 
magnetic and electric fields.
In that work, he predicted that a small electric field of the order of
1\,V\,cm$^{-1}$, even if contributing negligible linear polarization 
via the Stark effect, can still bring important modifications 
to the 
atomic polarization of the hydrogen levels when acting in the presence 
of a magnetic field. In particular, the simultaneous presence of 
magnetic and electric fields brings an enhancement of the A-O mechanism 
that is responsible for the appearance of net circular polarization
(NCP).


Figure \ref{fig.obs} does not show a significant amount of NCP. Since it 
is not possible to conclude from the data 
that electric fields may be producing any effect, we carried out 
the inversion of the Stokes spectra taking into account only the effect 
of magnetic fields.
We then estimated an upper limit for the electric field in the plasma 
by calculating the polarization degree in the simultaneous presence of 
magnetic and electric fields, assuming for the magnetic field the 
configuration derived from the inversion.

\subsection{Line formation in the presence of magnetic fields: PCA-based
inversion of the observations}
\label{sec.res.scatter_pca}

The inversion of the full Stokes profiles of the observed Paschen \ion{H}{1} 
lines in the presence of a magnetic field made use of pattern
recognition techniques. Such type of inversion involves the search 
of a precomputed database of Stokes profiles for the best match to 
the observed profiles.   
Specifically, we adopted Principal Component Analysis (PCA) for our 
implementation of 
pattern recognition inversion \citep{rees00, socas01, ariste02}. 

The code computing the inversion database of Stokes profiles solves the 
statistical equilibrium of a quantum model of the \ion{H}{1} atom in the 
presence of the magnetic field \citep{ariste02,landi04}, and computes 
from this solution the full Stokes vector of the scattered radiation. 
Figure~\ref{fig.geometry_code} shows the geometry of scattering in the presence of the magnetic field, and the parameters used in the line 
formation code.

For a good inversion, the database must cover the full range of parameters 
spanning the line formation model.
The parameters used to construct the precomputed database are the 
magnetic field strength, $B$, the inclination of the magnetic field 
vector with respect to the local solar vertical, $\vartheta_{B}$, 
the azimuth angle about the local solar vertical with respect to the 
$x'$ axis, $\varphi_{B}$, the 
inclination of the line-of-sight (LOS) from the local solar vertical, 
$\vartheta$, the height of the scattering point from the solar surface
along the local solar vertical, $h$, the optical thickness of the slab
at line center, $\tau$, and the temperature of the plasma in the
scattering region, $T$ (Fig. \ref{fig.geometry_code}).
The adopted range of these seven parameters, are 
$0.01<h<0.08\,R_\odot$, $0<B<1000$\,G, 
$0^\circ<\vartheta_B<180^{\circ}$, $0^\circ<\varphi_B<360^{\circ}$,
$82^\circ<\vartheta<98^{\circ}$,\footnote{We note that the 
LOS corresponding to the minimum height of $0.01\,R_\odot$ would 
intersect the solar disk if $\vartheta$ were outside this range.}
$0.01<\tau<1$, and $1000<T<65000$\,K.

One of the assumptions of the model is that the incident radiation on
the scattering atom is not polarized, and possesses cylindrical symmetry 
around the local solar vertical through the scatterer. In the
approximation of complete redistribution under which the line formation
model for this problem is valid, we must also assume that the radiation 
is spectrally flat over the width of the spectral line. 
Another assumption is to neglect collisional coupling among the atomic levels.
In Sect. \ref{sec.dis_sum} we discuss how this approximation affects the inference of
the magnetic and electric fields in the plasma.

The emergent Stokes parameters are obtained by solving the radiative 
transfer equation
\begin{equation}
	\frac{d}{ds}
	\left(
		\begin{array}{c}
		I  \\
		Q  \\
		U  \\
		V  \\
		\end{array}
	\right)
   = 
	\left(
		\begin{array}{cccc}
		\eta_{I}	&	\eta_{Q}	&	\eta_{U}	&	\eta_{V}		\\
		\eta_{Q}	&	\eta_{I}	&	\rho_{V}	&	-\rho_{U}	\\
		\eta_{U}	&	-\rho_{V}&	\eta_{I}	&	\rho_{Q}		\\
		\eta_{V}	&	\rho_{U}	&	-\rho_{Q}&	\eta_{I}		\\
		\end{array}
	\right)
	\left(
		\begin{array}{c}
		I  \\
		Q  \\
		U  \\
		V  \\
		\end{array}
	\right)
	+
	\left(
		\begin{array}{c}
		\epsilon_{I}  \\
		\epsilon_{Q}  \\
		\epsilon_{U}  \\
		\epsilon_{V}  \\
		\end{array}
	\right)
	,
	\label{eq.gen_rt}
\end{equation}
where $s$ is the geometrical distance along the ray passing through a 
slab and $(\eta_I, \eta_Q, \eta_U, \eta_V)$, 
$(\rho_Q, \rho_U, \rho_V)$, and 
$(\epsilon_I, \epsilon_Q, \epsilon_U, \epsilon_V)$ 
represent, respectively, the dichroic absorption, 
dispersion, and emission
coefficients for the Stokes parameters $I$, $Q$, $U$, and $V$.
The expression for each of those quantities is given by equations 
(7.47) of \citet{landi04}.
We assume homogenous physical conditions along the optical depth within the 
slab.
When the anomalous dispersion and dichroism terms are small, 
$\eta_{I} \gg (\eta_{X}, \rho_{X})$ (with $X=Q, U, V$), and the 
polarized emission is weak, $\epsilon_{I} \gg \epsilon_{X}$, we can formally integrate equation (\ref{eq.gen_rt}), yielding \citep{bueno05}
\begin{eqnarray*}
	I(\tau) & = & I_{0} e^{-\tau} + \frac{\epsilon_{I}}{\eta_{I}}(1-e^{-\tau})	\\
	X(\tau) & = & \frac{\epsilon_{X}}{\eta_{I}} (1-e^{-\tau}) 
		- I_{0} \frac{\eta_{X}}{\eta_{I}} \tau e^{-\tau}
		- \frac{\epsilon_{I} \eta_{X}}{\eta_{I}^{2}} [1-e^{-\tau}(1+\tau)]
	,
	\nonumber
\end{eqnarray*}
where $\tau$ (with $d{\tau}=-\eta_{I}ds$) is the optical depth along the 
ray, and $I_{0}$ is the background intensity of the slab (which 
is zero outside the limb).

The above approximate formulas for the emergent Stokes parameters from
a homogeneous and weakly polarizing slab are reasonable for the modeling
of solar chromospheric structures \citep{bueno05,bueno07}.

The adopted atomic model of \ion{H}{1} is composed of the terms in L-S 
coupling of principal quantum numbers $n=1,2,3,4$ plus the value of $n$ 
pertaining to the upper level of the observed transition.
L-S coupling is a reasonably good approximation for light atoms 
such as \ion{H}{1}. 
The reason for including the $n=4$ level in the
atomic model is because of the importance of the H$\alpha$ and H$\beta$ 
radiation in the solar spectrum in determining the polarization of the
$n=2,3,4$ levels.
The omission of the intermediate terms between $n=4$ and the upper level 
of the observed transition is justified because the statistical equilibrium 
of the upper level must be dominated by the optical pumping
through the resonant 
radiation at the frequency of the observed line from the $n=3$ level, as 
indicated by the predominance of the atomic polarization signature in
the emitted radiation. 

Figure \ref{fig.ali_ori} shows the estimated polarization degree 
for Stokes $Q$, $U$, and $V$, of the lines P7, P9, P10, and P11, 
for the case of $90^\circ$-scattering of the solar radiation at the limb. 
The polarization degree is calculated through $Q/I_\mathrm{max}$ and 
$U/I_\mathrm{max}$ at line center in the case of linear polarization, and
through the NCP in the case of Stokes $V$, as a function of the magnetic 
field strength, and for two geometric configurations of the magnetic
field.
In the first configuration (upper panel), the magnetic field is directed
along the LOS. For zero magnetic field, the scattered light is 
linearly polarized with positive $Q$ (i.e., parallel to the limb)  
as expected.
For $10^{-2} \,\mathrm{G} < B < 10 \,\mathrm{G}$, the direction of the 
linear polarization is rotated with respect to the zero-field case, and 
the degree of polarization is also decreased approaching zero around
10\,G. These rotation and reduction of the linear polarization of
scattered radition are characteristic phenomena of the magnetic Hanle
effect. For $B>1\,\mathrm{G}$, the atomic energy levels within a fine structure
term in the upper level start crossing each other, creating the
conditions for the Paschen-Back effect and the A-O mechanism to
generate asymmetric components in the Stokes $V$ profile as a function
of wavelength. 

In the second magnetic configuration (lower panel), the magnetic field 
is perpendicular to the LOS and parallel to the limb.  
For $10^{-2} \,\mathrm{G} < B < 10 \,\mathrm{G}$, the Hanle effect 
only depolarizes the scattered radiation, to about 1/2 of the
zero-field value. 
For $B$ larger than a few $10\,\mathrm{G}$, the generation of linear 
polarization by anisotropic pumping of radiation becomes more efficient 
due to the change of energy configuration of the atom in the transition 
to the Paschen-Back regime \citep{landi04,casini06}, when the spin-orbit
interaction becomes negligible with respect to the magnetic interaction.

The fundamental concepts of PCA inversion of Stokes profiles were described 
by \citet{rees00}. Applications of this technique to scattering polarization 
on the Sun were considered by \citet{ariste02,casini03,casini05a,casini09}.  
PCA is mainly a technique for pattern recognition and data compression. It
allows an efficient decomposition of the Stokes profiles using only a
very small number of ``universal'' principal components (eigenprofiles) 
out of the full basis that would be needed for a lossless reconstruction
of the profiles. 
The dimension of the PCA basis equals the number of wavelength points in 
the profiles. Since the number of principal components needed to 
reconstruct the Stokes profile to within the polarization sensitivity of
the observations is much smaller (typically by at least one order of
magnitude) than the number of wavelength points, PCA allows a much
faster comparison of the observed data with the model than iterative fitting.
Additionally, the use of a precomputed database of profiles for data
inversion is free from the risk of converging to local minima typical of
iterative fitting.

\subsection{Test of the inversion code}
\label{sec.res.test}

We examined the reliability of our inversion scheme by performing test 
runs for 10,000 synthetic Stokes profiles calculated with randomly 
distributed parameters in the 7D parameter space within the
ranges specified earlier, after adding to the
profiles a random noise of $3\times10^{-3}$ with respect to the peak 
intensity.
The inversion tests were run for P7, P9, P10, and P11, using a database 
of 100,000 models.
Figure \ref{fig.inverr} shows the inversion results against the 
actual values of the synthetic model for each parameter, for the case of
the P7 line.
The results for the inversion of the other lines are similar to the case 
of P7, and therefore we do not shown them here.

Figure \ref{fig.inverr}b shows the scatter plot of the longitudinal 
magnetic field strength ($B\,\cos{\Theta_{B}}$).
This quantity is determined rather well from the inversion of
Stokes $V$, which is mostly affected by the longitudinal Zeeman effect.
In contrast, the error in the inversion of the magnetic field 
strength and the inclination angle of the magnetic field vector is 
rather large (Figs.~\ref{fig.inverr}a and \ref{fig.inverr}c).
This can be understood when we consider that, for these Paschen lines, 
the Hanle effect is completely saturated already for 
$B > 10 \,\mathrm{G}$ (Fig.~\ref{fig.ali_ori}), whereas the Zeeman
effect does not produce a significant linear polarization until
the field reaches strengths of the order of a few kG. Therefore, for
$10\lesssim B\lesssim 1000$\,G the inferred magnetic field
strength is derived mainly from the circular polarization
signal.

Since there can be different values of the magnetic field strength
corresponding to a given degree of NCP or 
to the amplitude of the linear polarization induced by anisotropic
excitation (Fig. \ref{fig.ali_ori}), and considering that the atomic 
alignment from 
which both these polarization signals originate depends on the true 
height (versus the observed projected height) of the scatterer over the 
solar surface, as well as on the inclination of the magnetic field from
the local solar vertical $\vartheta_B$ (see eqs.~[\ref{eq.epcilon_qu}] below), 
we can understand why the inference of the magnetic
field strength and inclination can be affected by larger errors.


The scatter plot of the inverted azimuth angle of the magnetic field 
in the reference frame of the observer, $\Phi_{B}$, exhibits multiple 
streaks, which are caused by intrinsic ambiguities of scattering line polarization with the geometry of the field 
(Fig.~\ref{fig.inverr}d). Typical examples are the well-known
$180^\circ$-ambiguity of the Zeeman effect, and the $90^\circ$-ambiguity
of the saturated regime of the Hanle effect
\citep[e.g.][]{house77,casini99,casini02,casini05a}. 
%
%
In order to see this, we consider the emissivity for Stokes $Q$ and $U$ 
in the saturated regime of the Hanle effect for a two-level atom with 
unpolarized lower level, and neglecting stimulated emission. This is
simply expressed by 
\begin{equation}
\left.
\begin{array}{c}
	\epsilon_{Q}  \sim  w (3 \cos^2{\vartheta_{B}} -1) \sin^2{\Theta_{B}} \cos{2\Phi_{B}} 	\\
	\epsilon_{U}  \sim  w (3 \cos^2{\vartheta_{B}} -1) \sin^2{\Theta_{B}} \sin{2\Phi_{B}} 	
\end{array}\right.
,
\label{eq.epcilon_qu}
\end{equation}
where $w$ is the anisotropy factor of the radiation field \citep{casini02}.
A sign change of Stokes ${\it Q}$ and ${\it U}$ due to a 
$90^\circ$ change of $\Phi_{B}$ can thus be compensated by a sign
change of the factor $(3\cos^2{\vartheta_{B}}-1) \sin^2{\Theta_{B}}$.
Therefore the condition for the presence of the $90^\circ$ ambiguity 
(in the optically thin case) is the existence of 
$\vartheta_{B}'$ and $\Theta_{B}'$ such that
\begin{equation}
	(3\cos^2{\vartheta_{B}}-1) \sin^2{\Theta_{B}}  =  - (3\cos^2{\vartheta_{B}'}-1) \sin^2{\Theta_{B}'}.	
	\label{eq.90amb}	
\end{equation}
Since the dependence of the Stokes $Q$ and $U$ polarization on the 
geometry of the field and the observer, as given by 
eqs.~(\ref{eq.epcilon_qu}), is modified in the optically thick case,
or for the case of a multi-level atom, we have checked 
this dependence numerically in the optically thick case relying on our line 
formation model.
As a result, we verified that the expected variations of the polarization 
degree of the Paschen series with optical thickness remain below the
noise level of the observations, and thus we can still rely on 
eqs.~(\ref{eq.epcilon_qu}) to provide an approximate description of the
azimuthal ambiguities even for our more general atomic and radiative
transfer models.  
In the inversion results (see Sect.~\ref{sec.res.invres}), we determine
whether the inferred configuration is subject to ambiguities, by 
checking whether the condition (\ref{eq.90amb}) can be realized for that 
configuration.

If we ignore the ambiguities shown in Fig.~\ref{fig.inverr}d, the 
90\% confidence interval for the inversion of $\Phi_{B}$ is 
$\pm 13^{\circ}$ for P7, P9, P10, and P11.

The temperature is inferred from the line width under the assumption 
that this is determined by thermal broadening. The error on 
$T$ is estimated to be $0.2\,T$, based on the scatter plot of 
Fig.~\ref{fig.inverr}e. 

The determination of the remaining parameters is not very accurate.
The optical depth deforms the line shape and slightly changes the
polarization degree of the line.
The height of the scatterer determines the dilution and anisotropy 
factors of the incident radiation field, which in turn 
determines the state 
of atomic polarization in the zero-field case.
However, a change of the height over the entire parameter range,
between $0.01$ and $0.08\,R_\odot$ causes only small 
changes in the polarization degree of the lines.
The same is true for the change of $\vartheta$ between 
$82^{\circ }$ and $98^{\circ}$.

\subsection{Inversion results}
\label{sec.res.invres}

The observed Stokes spectra of the P7, P9, P10, and P11 lines were
inverted using the PCA technique illustrated in 
Sect.~\ref{sec.res.scatter_pca}, in order to derive the magnetic and
plasma model of the emitting plasma in the observed active region's 
jets
and surge. We did not invert the Paschen lines above P11, because the 
signal-to-noise ratios of the polarization signals were too low. 

For the inversion, the possible effects of an electric field 
(e.g., of the Lorentz type) were neglected. The Doppler velocity was 
derived from the shift of the center of gravity of the Stokes $I$ 
profiles, using the wavelengths of the Paschen lines as
reported by 
\citet{kramida10}. 

\begin{figure}
\begin{center}
\includegraphics[angle=0,scale=1.0,width=120mm]{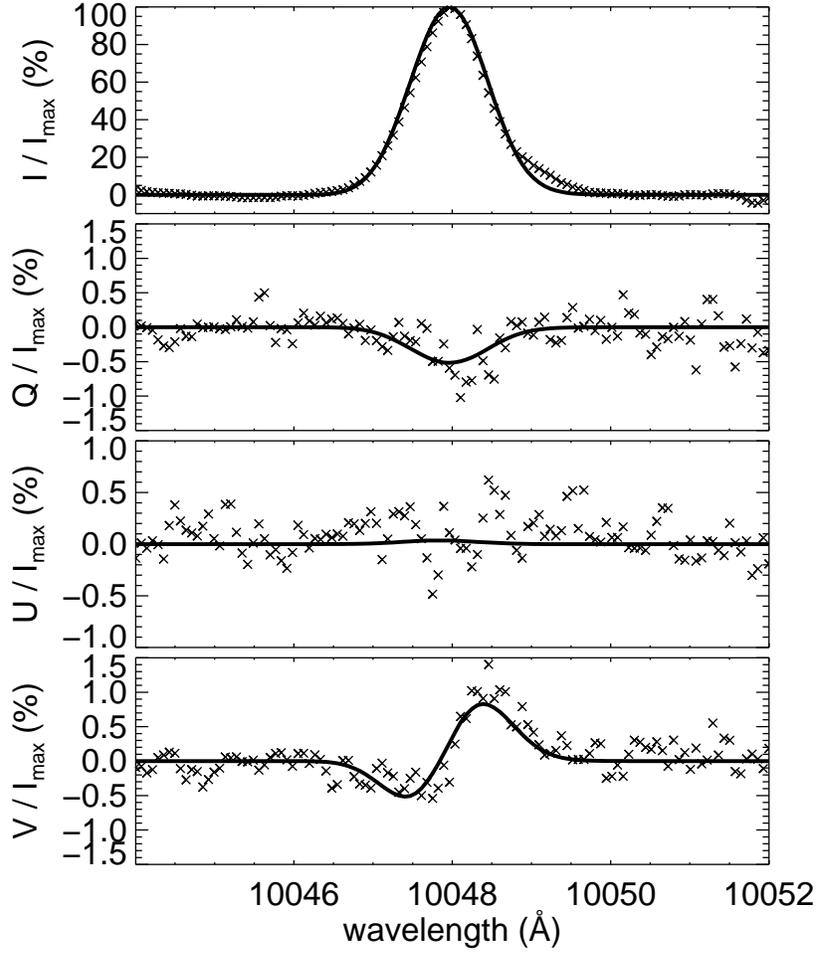}
\end{center}
\caption{
Example of the observed Stokes profiles (crosses) of the \ion{H}{1} P7 
line at 10049\AA\ and the inversion fits (solid curves).
The inverted magnetic field vector ($B,\vartheta_B,\varphi_B$) 
and temperature are approximately ($128\,\mathrm{G},112\mathrm{^{\circ}},
1.2\mathrm{^{\circ}})$ and $24300\,\mathrm{K}$, respectively.
			}
\label{fig.fit}
\end{figure}

\begin{figure}
\begin{center}
\includegraphics[angle=0,scale=1.0,width=100mm]{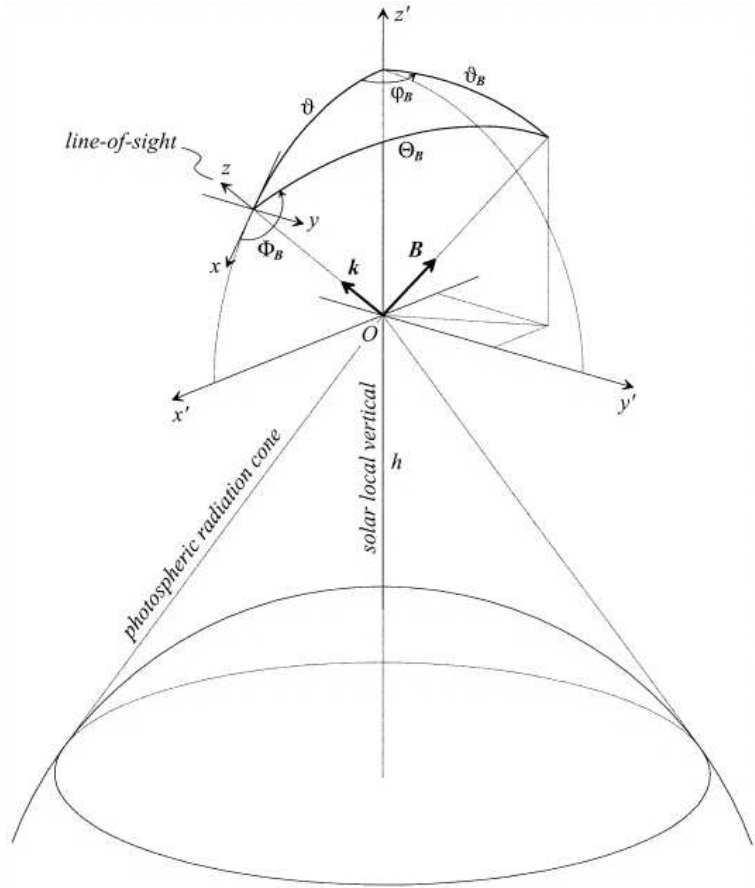}
\end{center}
\caption{
Geometric model for radiation scattering in the presence of a magnetic field. 
The cone of (photospheric) radiation irradiates the scattering atoms at 
the point $O$ of height $h$ above the solar surface.
The direction of the magnetic field vector, $\mathbf{B}$, is defined 
by the angles $\vartheta_B$ and $\varphi_B$ in the reference frame of 
the local solar vertical through the point $O$, and by $\Theta_B$ and 
$\Phi_B$ in the reference frame of the LOS, $\mathbf{k}$.
The inclination angle of $\mathbf{k}$ from the local solar vertical is 
$\vartheta$ and the direction of $y\equiv y'$, which corresponds to
the reference direction of positive Stokes $Q$, is parallel to the 
solar limb.
		}
\label{fig.geometry_code}
\end{figure}

\begin{figure}
\begin{center}
\includegraphics[angle=0,scale=1.0,width=160mm]{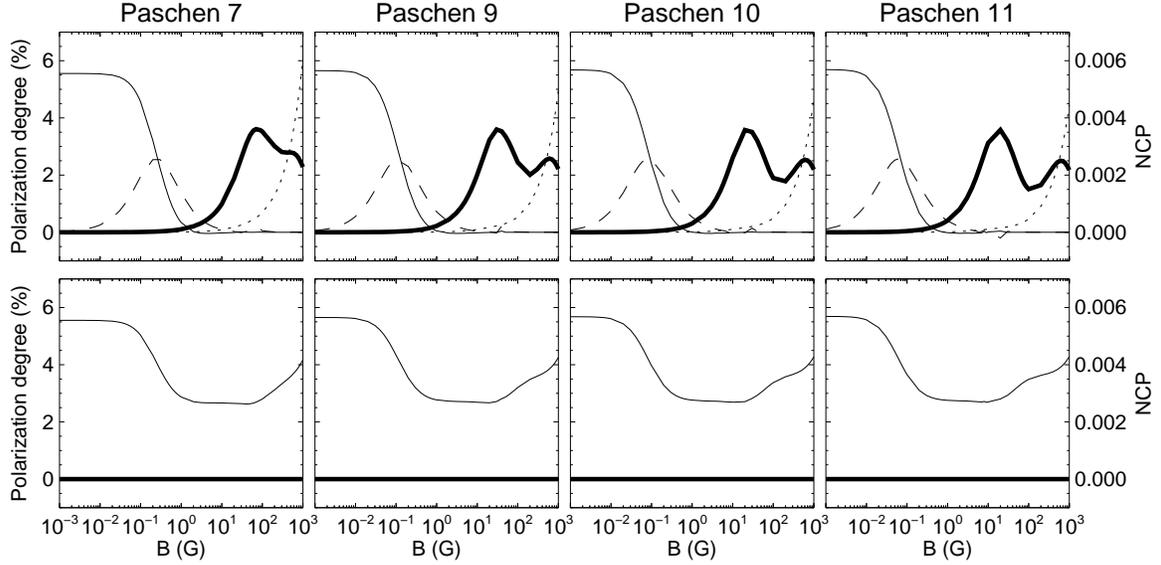}
\end{center}
\caption{
Expected polarization as a function of the magnetic field strength in 
the P7, P9, P10, and P11 lines, for $90^\circ$ scattering.
The top panels show solutions where the magnetic field is directed
along the LOS. The bottom panels show solutions were
the magnetic field lies in the POS, and 
is perpendicular to the solar vertical (i.e., horizontal; 
$\mathbf{B}$ along the $y$-axis of Fig.~\ref{fig.geometry_code}).
The thin solid lines and the dashed lines show, respectively, 
$Q/I_\mathrm{max}$ and $U/I_\mathrm{max}$ at line 
center. The dotted lines show $V/I_\mathrm{max}$ at $+0.5$\AA\ from 
line center. The thick solid lines show the NCP.	
		}
\label{fig.ali_ori}
\end{figure}

\begin{figure}
\begin{center}
\includegraphics[angle=0,scale=1.0,width=160mm]{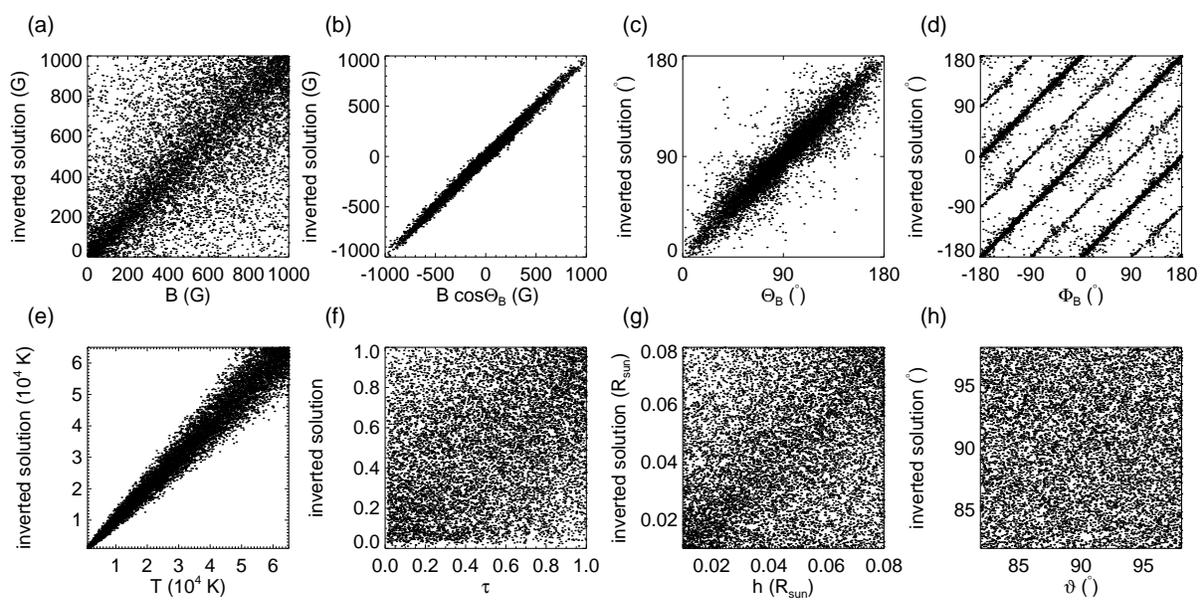}
\end{center}
\caption{
Scatter plots of the inverted vs true values of the model parameters
for 10,000 random synthetic profiles of P7. The PCA-based inversions 
used a database of 100,000 models spanning the same parameter ranges 
as the synthetic profiles.
		}
\label{fig.inverr}
\end{figure}

Figure \ref{fig.fit} shows an example of inversion fit of the Stokes 
vector of P7 for a jet, the location of which is indicated by the dashed 
lines in the top of Fig.~\ref{fig.obs}.
Our spectral synthesis does not reproduce the complex shape of Stokes profiles 
produced in a spatially unresolved, multi-component atmosphere, because of 
the assumption of a homogeneous slab.
Nonetheless, most of the observed Stokes profiles fit satisfactorily 
some synthetic profile in our inversion database with 
only 100,000 models, as shown in Fig.~\ref{fig.fit}.



The distributions of the inverted parameters along the slit are shown in 
Figure~\ref{fig.scatres}.
The ranges of the physical parameters estimated from the inversion
corresponding to the 90\% confidence level (Sec. \ref{sec.res.test}) are
shown in the four bottom rows, respectively for the longitudinal component 
of the magnetic field, the magnetic field strength, the plasma temperature, 
and the plasma velocity along the LOS.
In the top row, the direction of the lines indicates the direction of the 
magnetic field on the POS (related to $\Phi_B$). 
The red (blue) color marks the solutions which admit (do not admit) an 
alternative
$90^\circ$-ambiguous configuration of the magnetic field.
The magnetic field solutions that do not admit this azimuthal 
ambiguity indicate that the projected magnetic field on the POS
is approximately aligned to the jets and the surge.
Accordingly, 60\% of the solutions marked with the red lines were
rotated by 90$^\circ$ in order to align the projected magnetic field on 
the POS to the directions of the jets and the surge.
After this operation, it is noticeable that the direction 
of the magnetic field on the POS appears to be 
systematically tilted away from the surge by about $25^{\circ}$
counterclockwise, which is larger than the estimated inversion error 
of $13^{\circ}$ of $\Phi_{B}$ corresponding to the 90\% confidence 
level. We do not speculate on the physical origin of this tilt.

The inverted magnetic field component along the LOS 
($B_\parallel=B\cos{\Theta_{B}}$) is comprised in the following ranges,
respectively for the three jets and the surge:
$0\,\mathrm{G} < B_\parallel < 180 \,\mathrm{G}$, 
$80\,\mathrm{G} < B_\parallel < 250 \,\mathrm{G}$, 
$10\,\mathrm{G} < B_\parallel < 160 \,\mathrm{G}$, and 
$-70\,\mathrm{G} < B_\parallel < 100 \,\mathrm{G}$.

For the magnetic field strength of the three jets, using the
observation at 2:47 UT, we found
$60\,\mathrm{G} < B < 630 \,\mathrm{G}$, 
$190\,\mathrm{G} < B < 960 \,\mathrm{G}$, and 
$40\,\mathrm{G} < B < 200 \,\mathrm{G}$.
In the case of the surge, the circular polarization is practically
absent, leading to a larger uncertainty in the determination of the
magnetic field strength in that structure. For a 90\% confidence level 
we found $10\,\mathrm{G} < B < 640 \,\mathrm{G}$.

Figure \ref{fig.scatres} shows other parameters that are determined by 
the inversion. The derived temperature must be regarded as an upper 
limit, because only thermal and natural broadening were considered for
the inversion, while other broadening mechanisms such as pressure
broadening, plasma turbulence, and collisional damping may also
contribute to the effective line width.
The maximum Doppler velocity of the surge at 3:11 UT, 3:13 UT, 
and 3:16 UT was $7\,\mathrm{km\,s^{-1}}$, $4\,\mathrm{km\,s^{-1}}$, 
and $16\,\mathrm{km\,s^{-1}}$, respectively.

\subsection{Upper limit of the electric field}
\label{sec.res.elec}

The time evolution of the surge can be seen in Figure~\ref{fig.v}, 
which shows a slit-jaw image at the line core of 
H$\alpha$ (left) and 
two time-distance diagrams showing the motion along two different 
sections of the surge, respectively parallel (top right) and 
perpendicular (bottom right) to it.

The surge first appeared at 2:58 UT, ejected at a speed of 
$\sim 100\,\mathrm{km\,s^{-1}}$, initially with a whiplike 
motion, finally reaching a length of $\sim 110,000 \,\mathrm{km}$.
The observations taken at 3:11 UT, 3:13 UT, and 3:16 UT
show the surge near its maximum extension.
The observed Doppler shift indicates a significant component of 
plasma velocity along the LOS, which is also nearly 
perpendicular to the magnetic field, if we accept the inversion 
results that the magnetic field vector practically lies in
the POS (Fig.\ref{fig.scatres}).
Therefore, this event provides a good opportunity for testing the 
coupling between partially ionized plasma and magnetic fields, 
since neutral atoms that move across the magnetic field must 
experience a motional electric field.

We used the inverted magnetic and velocity field configurations of 
the surge as the basis to calculate the emergent Stokes profiles in 
the additional presence of an electric field (Fig. \ref{fig.ponti}).
The magnetic field, which is roughly aligned to the surge, is inclined 
about $45^{\circ}$ from the local solar vertical, and on the 
POS (Fig.~\ref{fig.scatres}). The velocity 
component is assumed
to be along the LOS, because no significant apparent 
lateral motions were observed in the H$\alpha$ slit-jaw images at the 
times of the observation, 3:11 UT, 3:13 UT, and 3:16 UT (Fig. \ref{fig.v}).
Since the inferred magnetic field strength in the surge
varies approximately between
$10\,\mathrm{G}<B<640\,\mathrm{G}$, we calculated the 
expected line polarizations in the presence of a motional electric field 
for three magnetic-field strengths of $70$, $200$, and 
$600\,\mathrm{G}$.
Unlike the surge, the magnetic field in the jets has a significant
component along the LOS. Since it is difficult to estimate 
the velocity component across the magnetic field from the time
evolution of the plasma in the observations, we omit any discussion 
about the estimation of electric fields in the jets.

We calculated the theoretical net linear and circular polarizations of 
the Paschen \ion{H}{1} lines as functions of the strength of the
motional electric field, assumed to be perpendicular to both the 
magnetic field and the LOS, using the formalism of 
\citet{casini05b}.
Because of the particular field geometry (both fields
lie on the POS), all configurations where the orientation of
either of the two fields is inverted give rise to the same polarization.

Figure \ref{fig.e_pd} shows the results for P10 for the three selected
values of the magnetic field strength.
It is notable that the direction of the linear polarization does not 
change appreciably when the electric field is applied. This is
in fact dominated by Stokes $U$ at all times, in agreement with the fact
that the atom is in the saturated regime of the (magnetic) Hanle effect,
and therefore the direction of linear polarization
must be perpendicular to the magnetic field direction 
($135^\circ$ in this calculation).
This fact allows us to determine the azimuth angle of the magnetic 
field ($\Phi_{B}$) from the inversion independently of the effect of 
possible electric fields.

The upper limit of the electric field in the P10 at 3:13 UT is estimated 
by comparing the observed polarization degree with the 
calculated
values. Because of the residual continuum polarization observed 
in the Stokes spectra after correcting them for instrumental polarization
(see end of Sect.~2), the observed polarization degree
is affected by both random and systematic errors. 
In Figure~\ref{fig.e_pd}, the horizontal dotted lines in each
plot limit the range of polarization error due to both systematic
(either instrumental stray light or residual polarization cross-talk) 
and random sources. Additional polarization effects due to an undetected 
motional electric field must then lie within this range.

We thus find an upper limit of $0.04$, $0.3$, and 
$0.8\,\mathrm{V\,cm^{-1}}$, respectively for the magnetic field 
strength of $70$, $200$, and $600\,\mathrm{G}$. 
This implies that the upper limit of the velocity of neutral hydrogen 
moving across the magnetic field corresponds respectively to 
$0.6$, $1.5$, and $1.3\,\mathrm{km\,s^{-1}}$. 
Since the measured Doppler velocity in the P10 at 3:13 UT is 
$3.4\pm1.6\,\mathrm{km\,s^{-1}}$, the velocity of neutral hydrogen 
moving across the magnetic field appears to be significantly 
smaller than the plasma's bulk velocity.

In the same way, we estimated the upper limit of the velocity of neutral 
hydrogen moving across the magnetic field at 3:11 UT from observations
of the P9 line. We found in this case the values $0.8$, $1.5$, and 
$1.1\,\mathrm{km\,s^{-1}}$ for the same magnetic field strengths,
in very good agreement with the results found with P10.
Using the P11 data (observed at 3:16 UT) we found instead $1.1$, 
$1.0$, and $0.8\,\mathrm{km\,s^{-1}}$, which also are in 
agreement with the results from P9 and P10.
Since the measured Doppler velocity in the P9 at 3:11 UT, and in 
the P11 at 3:16 UT, are $3.3\pm0.2\,\mathrm{km\,s^{-1}}$ and 
$12.5\pm1.4\,\mathrm{km\,s^{-1}}$, respectively, all estimated upper 
limits of the velocity across the magnetic field are smaller than the 
measured Doppler velocities.

\section{Discussion and summary}
\label{sec.dis_sum}

We presented magnetic field inferences in a surge and in active region 
jets, and a first estimation of the upper limit of motional electric
fields, and the corresponding limit to plasma velocities across the
magnetic field lines.
 
The direction of the magnetic field on the POS is found 
to approximately align to the jets and the surge.
This confirms the common scenario of the chromospheric jet model, in which 
plasma is ejected along the magnetic field \citep[e.g.][]{shibata07,takasao13}.  
When we look carefully at the geometric relation between the magnetic field and the 
jet, we find that the direction of the magnetic field is slightly tilted 
counterclockwise from the direction of the surge.
Further study with high-accuracy spectro-polarimetric data for a number 
of similar events is needed in order to clarify the details of the 
geometric relation between the magnetic field and jets.

The strength of the magnetic field in the three jets is found to be in
the ranges $60\,\mathrm{G} < B < 630 \,\mathrm{G}$, 
$190\,\mathrm{G} < B < 960 \,\mathrm{G}$, and 
$40\,\mathrm{G} < B < 200 \,\mathrm{G}$.
This determination is driven mainly by the observed 
circular polarization signal and its symmetry characteristics.
Because NCP can also be generated by the coupling of 
velocity and magnetic field gradients along the LOS \citep{illing74},
and since our model does not contemplate the possibility of a multi-component 
atmosphere, we cannot exclude the possibility that the inferred magnetic field 
strengths may be affected by the possible presence of velocity and
magnetic field inhomogeneities along the LOS.

The upper limit of the motional electric field is estimated to be 
$0.04\,\mathrm{V\,cm^{-1}}$ for a magnetic field strength of 
$70\,\mathrm{G}$, based on the observed polarization degree in the P10 
line.
If we neglect the magnetic field and the atomic level polarization,
as assumed in previous estimates of solar electric fields, the 
limit to the electric field that is derived from the observed 
polarization degree (using, e.g., the formalism of \citealt{goto08}) is 
$\sim 30\,\mathrm{V\,cm^{-1}}$, which is significantly larger than
our estimates.  

In our inversion model, we neglected collisional coupling among the 
atomic levels, which tends to reduce the amount of linear polarization 
and NCP in the scattered radiation.
In the case of the surge, the observed linear polarization may be
explained through a combination of collisional depolarization with the
polarizing effects of the magnetic field and the possible motional
electric fields in the plasma (Fig.~\ref{fig.e_pd}).
In particular this implies
that our inferred upper bounds for the motional electric field present
in the surge could be underestimated. Similarly, our model
might underestimate the LOS component of the magnetic field, if the
observed low values of NCP were caused by the destruction of atomic
orientation by collisions. 
Finally, we note that the presence of
isotropic collisions does not cause a significant change of the direction of the emergent
linear polarization, hence we can still reliably determine the azimuth
angle of the magnetic field on the POS ($\Phi_{B}$) from the inversion,
independently of the effects of collisions.

\begin{figure}
\begin{center}
\includegraphics[angle=0,scale=1.0,width=160mm]{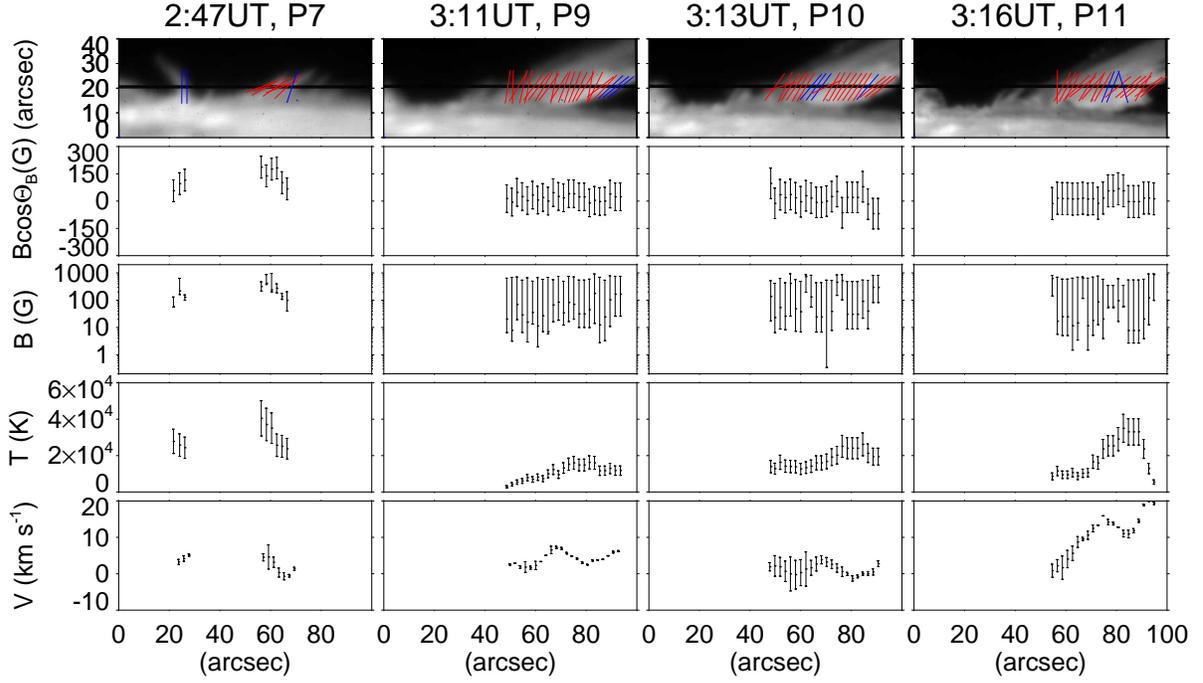}
\end{center}
\caption{
Inverted magnetic field, temperature, and Doppler velocity for the 
observed jets and surge at the position of the slit shown in the figure.
The columns, from left to right, show the inversion results 
for the observed Stokes spectra of P7 (2:47 UT), P9 (3:11 UT), P10 (3:13
UT), and P11 (3:16 UT).
The slit-jaw images in the top row show the direction of the magnetic 
field on the POS. The red lines indicate inverted solutions 
affected by the $90^\circ$ azimuthal ambiguity, while the blue lines 
correspond to the solutions with nonambiguous azimuth. 
The second to fifth rows show, respectively, the longitudinal 
component of the magnetic field, the magnetic field strength, 
the temperature, and the Doppler velocity.
The abscissas indicate the position along the slit.
		}
\label{fig.scatres}
\end{figure}

\begin{figure}
\begin{center}
\includegraphics[angle=0,scale=1.0,width=160mm]{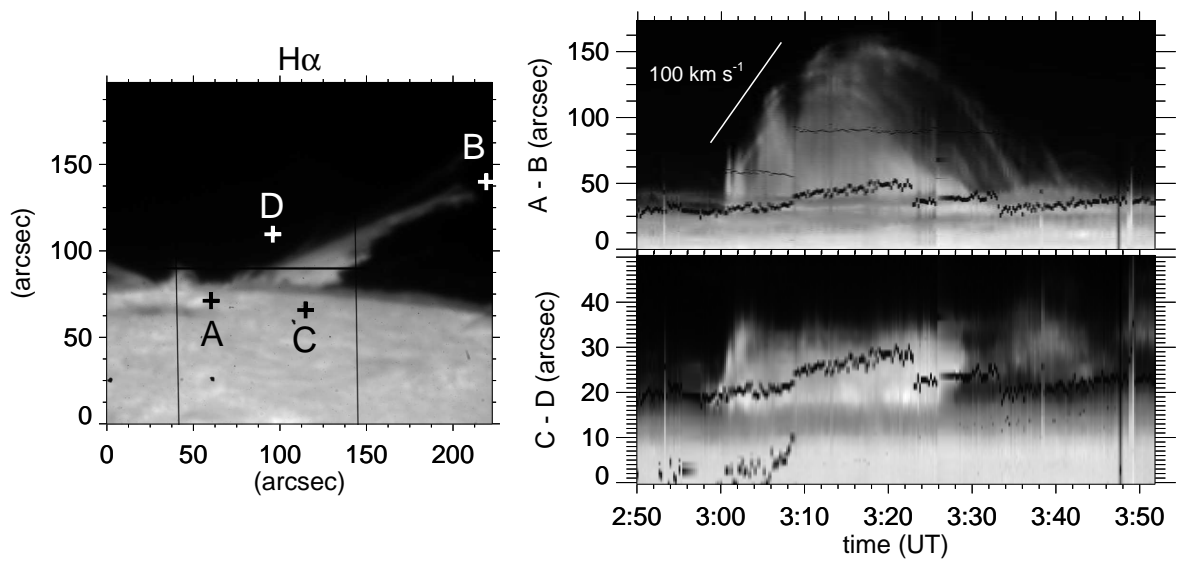}
\end{center}
\caption{
\emph{Left:} Slit-jaw image in H$\alpha$ of the surge observed at 3:16 UT.
\emph{Right:} Time-distance diagrams obtained between the points 
A and B (top), and C and D (bottom), marked on the slit-jaw image on the
left.
The thick dots in the diagrams show the spectrograph slit. The thin dots 
in the top diagram show one of the spectrograph hairlines.
		}
\label{fig.v}
\end{figure}

\begin{figure}
\begin{center}
\includegraphics[angle=0,scale=1.0,width=100mm]{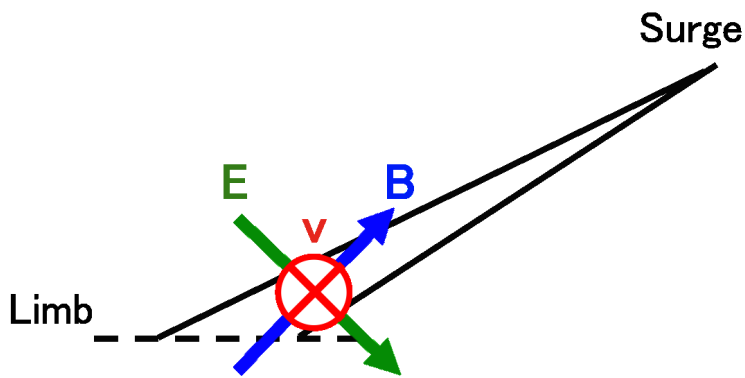}
\end{center}
\caption{
Schematic picture of the magnetic, electric, and velocity vectors in the
surge.
		}
\label{fig.ponti}
\end{figure}

\begin{figure}
\begin{center}
\includegraphics[angle=0,scale=1.0,width=100mm]{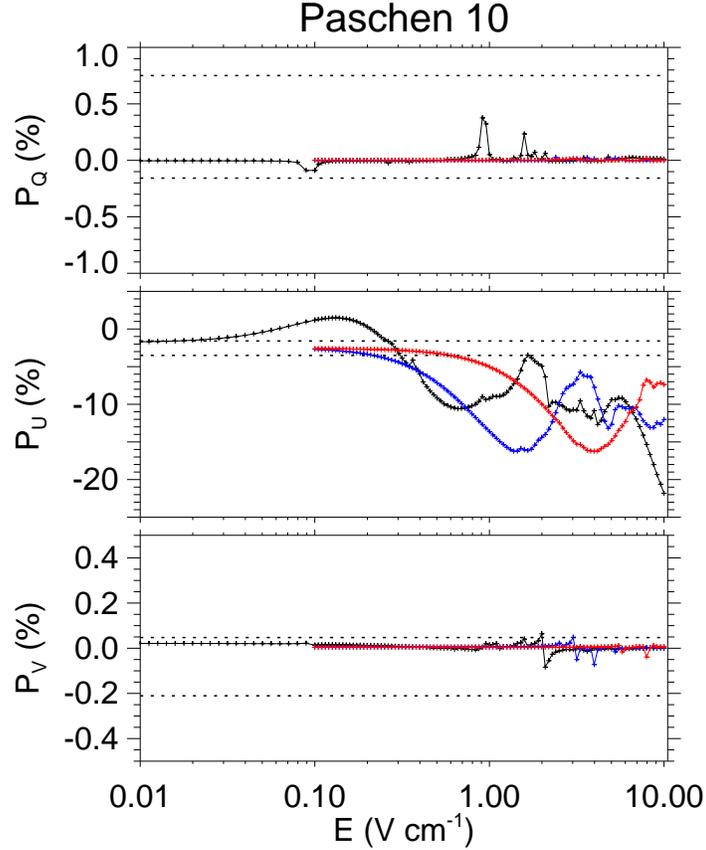}
\end{center}
\caption{
Broadband polarization of P10 in Stokes $Q$ (top), $U$ (center), and $V$
(bottom), as a function of the electric field strength, in a
$90^\circ$ scattering event. The magnetic field lies on the POS, with 
azimuth $\Phi_B=135^\circ$ (see Fig.~\ref{fig.geometry_code}).
The electric field also lies on the POS, and is 
assumed to be perpendicular to the magnetic field.
Black, blue, and red lines correspond to magnetic field strengths of $70$, 
$200$, and $600\,\mathrm{G}$, respectively.
The horizontal lines in each plot limit the range of polarization 
error due to both random and systematic (instrumental stray
light and residual polarization cross-talk) sources.
		}
\label{fig.e_pd}
\end{figure}

From the upper limit of the motional electric field, we can
estimate the maximum
velocity of neutral hydrogen moving across the ambient magnetic field.
At 3:16 UT, the surge reaches its maximum height (Fig.~\ref{fig.v}) 
and a large bulk velocity (i.e., Doppler shift) is observed 
(Fig. \ref{fig.scatres}). 
Therefore the velocity vector can be assumed to be along the LOS 
and perpendicular to the magnetic field (Fig.~\ref{fig.scatres}).
On the other hand, from our analysis of the polarization effects
of motional electric fields, the upper limit of the velocity of neutral 
hydrogen moving across the magnetic field is found to be 
$1.1\,\mathrm{km\,s^{-1}}$ in this 
geometry. As this is much smaller than the measured bulk
velocity of $12.5\pm1.4\,\mathrm{km\,s^{-1}}$, our estimate of the upper bound of the motional electric fields
in the plasma leads us to conclude that 
neutrals must be in a highly frozen-in condition in the surge.


\acknowledgments

This work was supported by a Grant-in-Aid for Scientific Research 
(No. 22244013, P.I. K. Ichimoto) from the Ministry of Education, 
Culture, Sports, Science and Technology of Japan, and by the 
Grant-in-Aid for the Global COE Program "The Next Generation of 
Physics, Spun from University and Emergence" from the Ministry of 
Education, Culture, Sports, Science and Technology (MEXT) of Japan.
This grant supported a two-month visit by T.A. at the High
Altitude Observatory (HAO) of the National Center for 
Atmospheric Research (Boulder, CO), where the interpretational work 
presented in this paper was initiated.
The authors acknowledge all the staff and students of Kwasan and 
Hida Observatory, especially Mr.\ S.~Ueno and Mr.\ A.~Oi who helped 
with the observations. 
The authors are also grateful to R.~Centeno Elliott (HAO) for a
careful reading of the manuscript and helpful comments.



\clearpage


\begin{deluxetable}{cccccccc}
\tabletypesize{\scriptsize}
\rotate
\tablecaption{Observation properties}
\tablewidth{0pt}
\tablehead{
\colhead{time (UT)} & \colhead{target} & \colhead{line} & \colhead{$\lambda$ (\AA)} & \colhead{noise$^a$}
}
\startdata
2:47 & jets 	& P7 	& 10049.368	& $2\times10^{-3}$	\\
3:11 & surge	& P9 	& 9229.014 	& $2\times10^{-3}$	\\
3:13 & surge	& P10 	& 9014.909	& $3\times10^{-3}$	\\
3:16 & surge	& P11 	& 8862.782	& $3\times10^{-3}$	\\
3:23 & surge	& P12 	& 8750.472	& $3\times10^{-3}$	\\
3:25 & surge	& P13 	& 8665.019	& $4\times10^{-3}$	\\
3:30 & surge	& P15 	& 8545.383	& -$^b$ \\
3:34 & surge	& P18	& 8437.956	& -$^b$ \\
3:34 & surge	& P19	& 8413.318	& -$^b$ \\
3:39 & surge	& P20 	& 8392.387	& -$^b$ \\
\enddata
\tablenotetext{a}{Noise with respects to the peak intensity of spectral line}
\tablenotetext{b}{Signal is too low to be detected.}
\end{deluxetable}

\end{document}